\documentclass[runningheads]{llncs}
\usepackage{graphicx}
\usepackage{hyperref}
\usepackage{amsmath}
\usepackage{tabu}
\usepackage{xcolor}
\usepackage{placeins}
\usepackage{subcaption}
\usepackage{graphicx}
\usepackage{bbding}

\begin{document}

\title{Reducing Textural Bias Improves Robustness of Deep Segmentation Models}
\author{Seoin Chai \and Daniel Rueckert  \and Ahmed E. Fetit~\textsuperscript{(}\Envelope\textsuperscript{)}
}
\authorrunning{S. Chai et al.}
\institute{Imperial College London, London, UK 
\\ 
\email{a.fetit@imperial.ac.uk}}
%

%
%
\maketitle              
\begin{abstract}
Despite advances in deep learning, robustness under domain shift remains a major bottleneck in medical imaging settings. Findings on natural images suggest that deep neural models can show a strong textural bias when carrying out image classification tasks. In this thorough empirical study, we draw inspiration from findings on natural images and investigate ways in which addressing the textural bias phenomenon could bring up the robustness of deep segmentation models when applied to three-dimensional (3D) medical data. To achieve this, publicly available MRI scans from the Developing Human Connectome Project are used to study ways in which simulating textural noise can help train robust models in a complex semantic segmentation task. 
We contribute an extensive empirical investigation consisting of 176  experiments and illustrate how applying specific types of simulated textural noise prior to training can lead to \textit{texture invariant} models, resulting in improved robustness when segmenting scans corrupted by previously unseen noise types and levels.

\end{abstract}
\section{Introduction}
In medical imaging research, using convolutional neural networks (CNNs) is a popular way of classifying and segmenting scans. However, medical scans can contain subtle visual noise and low-frequency textural patterns that are inherent to acquisition protocols or hardware. As a result, a network may be optimised to reduce empirical risk on data from one domain, but its performance can degrade when applied to a different domain, such as a different hospital or imaging centre \cite{kamnitsas2017,perone2019}. In a study on natural images by Geirhos et al. \cite{geirhos2018}, separate CNNs were trained on standard ImageNet \cite{deng2009imagenet} data as well as a stylised version of the same data, introducing conflicting textures to the input images. In doing so, the authors showed that training networks on stylised images has led to a reduction of the textural bias phenomenon as the models focused on more robust shape features. In this study, we draw inspiration from the work reported in \cite{geirhos2018} and hypothesise that addressing this phenomenon in medical imaging can lead to CNNs that are \textit{texture invariant} and hence more resilient to changes in data distribution. Specifically, our motivation is  to  find  out  whether  training segmentation CNNs  under certain  textural  noise  settings  could  help improve model robustness. Long term benefits of this include flexibility to variations in acquisition protocols, scanner hardware, hospital infrastructure, or image resolution often faced in realistic healthcare scenarios and routine clinical practice.

In this regard, we carried out an empirical investigation using neuroimaging data publicly available from the Developing Human Connectome Project (dHCP)\footnote{\url{http://www.developingconnectome.org/project/}} \cite{bastiani2018}. In doing so, we simulated different categories and levels of textural noise by applying several permutations of filtering techniques to the scans. In a series of 176 experiments, we trained 11 models and subsequently tested them on 16 different held-out sets in order to evaluate which settings can best generalise to previously unseen noise, thereby thoroughly simulating performance under various types and severities of domain shift.

Our contribution is an extensive empirical study which demonstrated that training a deep segmentation model on data corrupted with certain combinations of textural noise can in fact improve model robustness on new, previously unseen noise types and levels. We believe that this is due to the models being incentified to learn anatomical and tissue-specific features during training, as opposed to low-frequency textural patterns that may be brittle and domain specific. 

The rest of the paper is structured as follows: in Section \ref{Dataset} we present a summary of the dataset used, followed by an explanation of the three main textural noise simulation techniques used in Section \ref{noise}. Then, we give a summary of the CNN architecture used in Section \ref{cnn}, as well as a breakdown of the 176 experiments conducted in Section \ref{experiments}. We present and discuss details of the experimental results in Section \ref{results}, which includes tables of three insightful models used in our experiments, as well as visual examples of how the most robust model performs on previously unseen noise levels. Finally, we present conclusions and directions for future work in Section \ref{sec:conclusion}.

\section{Materials and Methods}

\subsection{Dataset} \label{Dataset}

\begin{figure}[h]
\begin{subfigure}[t]{.25\textwidth}
\centering
\includegraphics[width=.95\linewidth]{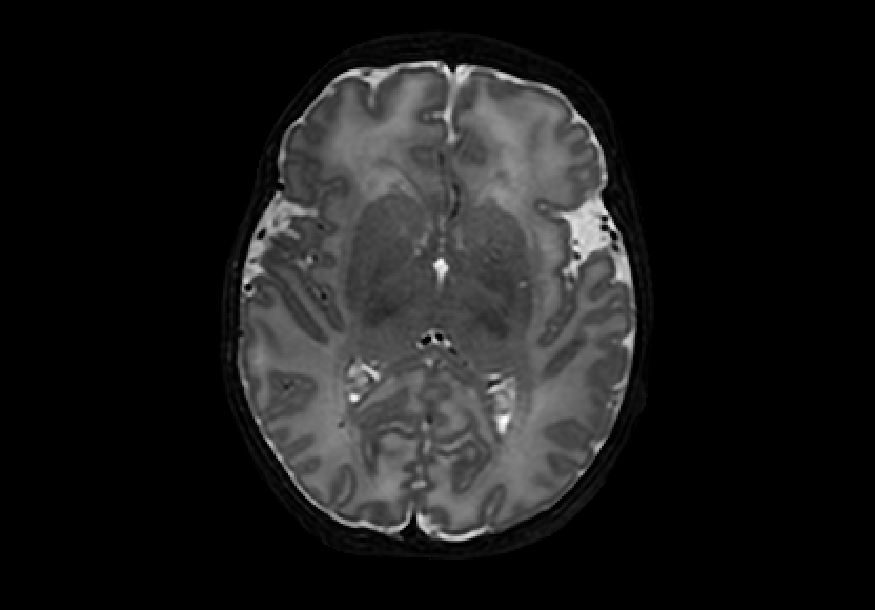}
\caption{T2 axial}
\label{subfig:dHCP_train_t2_axial}
\end{subfigure}%
\begin{subfigure}[t]{.25\textwidth}
\centering
\includegraphics[width=.95\linewidth]{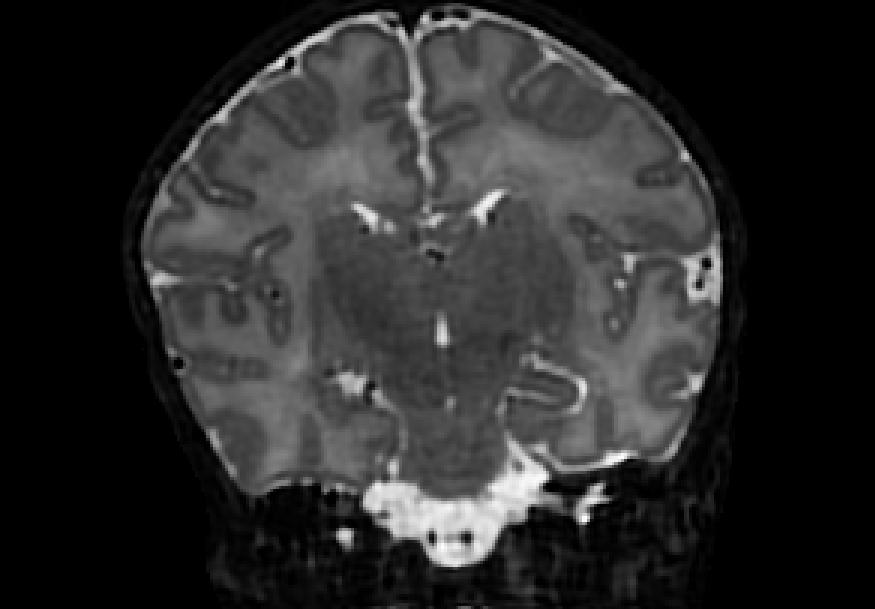}
\caption{T2 coronal}
\label{subfig:dHCP_train_t2_coronal}
\end{subfigure}%
\begin{subfigure}[t]{.25\textwidth}
\centering
\includegraphics[width=.95\linewidth]{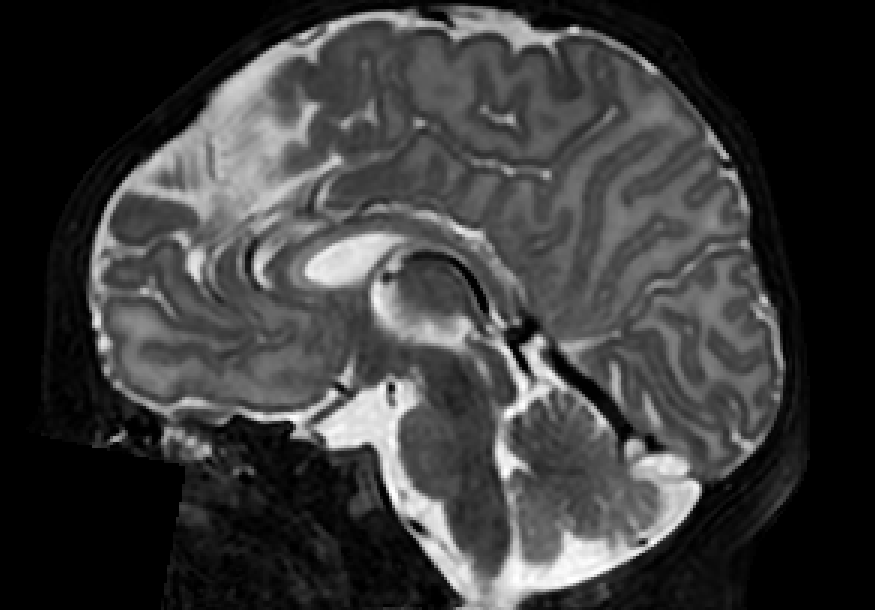}
\caption{T2 sagittal}
\label{subfig:dHCP_train_t2_sagittal}
\end{subfigure}%
\begin{subfigure}[t]{.25\textwidth}
\centering
\includegraphics[width=.95\linewidth]{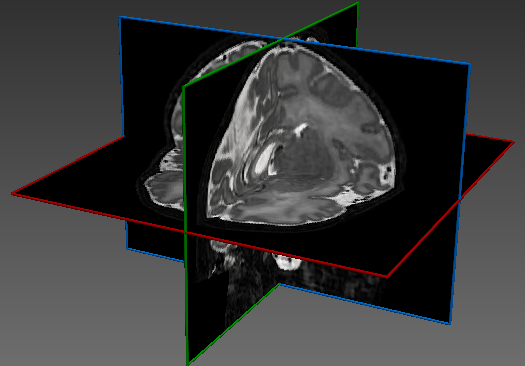}
\caption{T2 3D}
\label{subfig:dHCP_train_t2_3D}
\end{subfigure}

\medskip

\begin{subfigure}[t]{.25\textwidth}
\centering
\includegraphics[width=.95\linewidth]{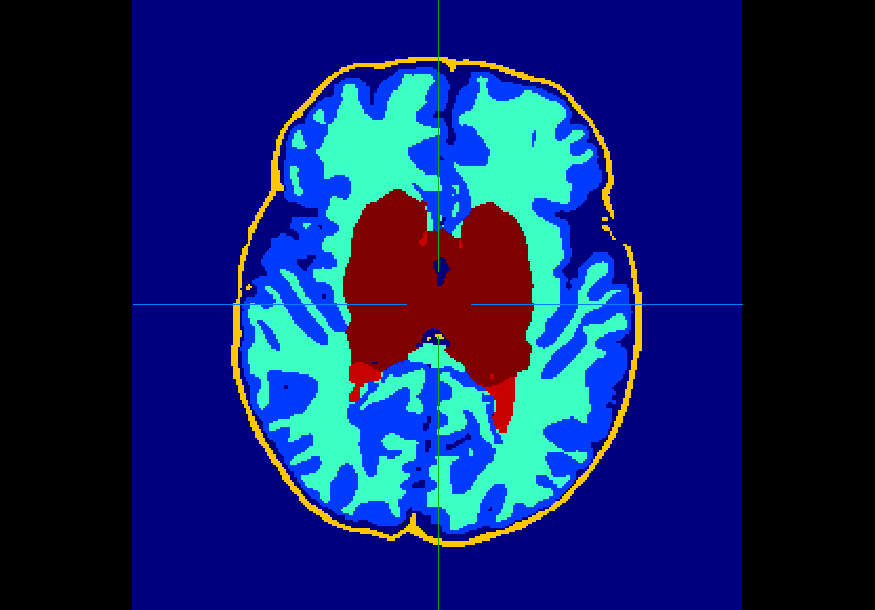}
\caption{Labels axial}
\label{subfig:dHCP_train_seg_axial}
\end{subfigure}%
\begin{subfigure}[t]{.25\textwidth}
\centering
\includegraphics[width=.95\linewidth]{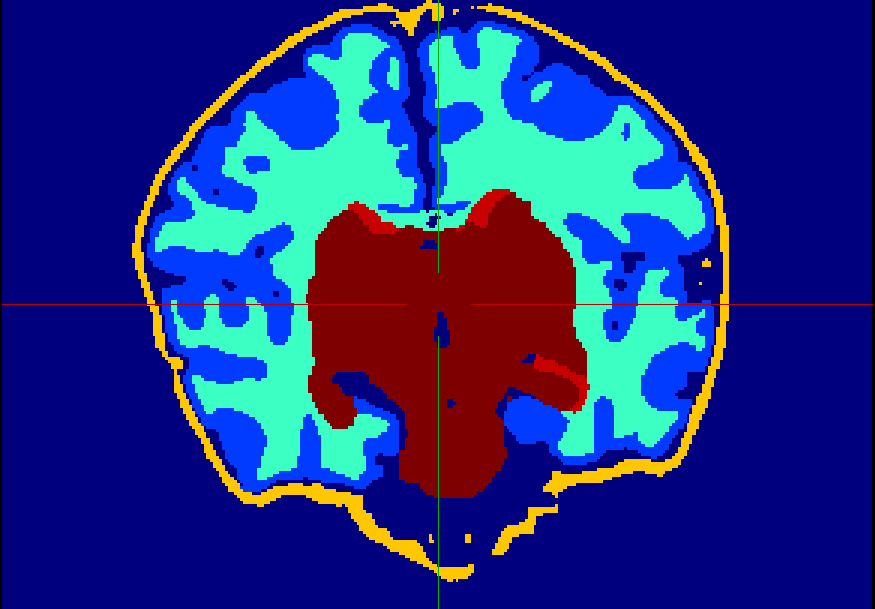}
\caption{Labels coronal}
\label{subfig:dHCP_train_seg_coronal}
\end{subfigure}%
\begin{subfigure}[t]{.25\textwidth}
\centering
\includegraphics[width=.95\linewidth]{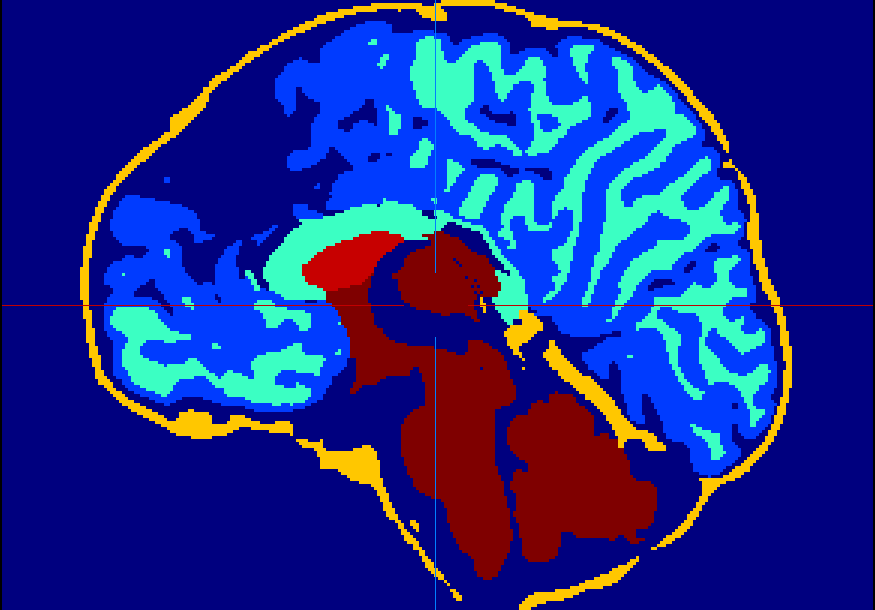}
\caption{Labels sagittal}
\label{subfig:dHCP_train_seg_sagittal}
\end{subfigure}%
\begin{subfigure}[t]{.25\textwidth}
\centering
\includegraphics[width=.95\linewidth]{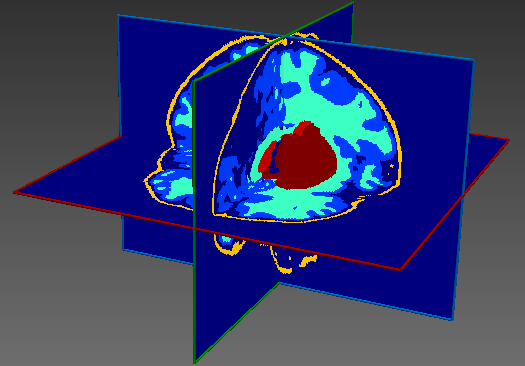}
\caption{Labels 3D}
\label{subfig:dHCP_train_seg_3D}
\end{subfigure}
\caption{Example T2-weighted neonatal brain scan and corresponding segmentation labels from the publicly available dHCP dataset.}\label{fig:dHCP_train_t2_seg}
\end{figure}

The dataset consisted of 70 3D T2-weighted brain MRI scans publicly available from the dHCP neonatal cohort. The segmentation maps had 10 classes, corresponding to: zero-pixel background, cerebrospinal fluid (CSF), cortical grey matter (cGM), white matter (WM), background bordering brain tissues, ventricles, cerebellum, deep grey matter (dGM), brainstem, and hippocampus. The scans covered an age range of 24.3-42.2 weeks. The data was available in NIfTI format; Figure~\ref{fig:dHCP_train_t2_seg} shows an example scan and corresponding tissue labels. We carried out a pre-processing step where each scan was independently normalised to zero-mean and unit-variance. 

\subsection{Textural Noise Simulation}\label{noise}

\begin{figure}[!htb]
\centering
\begin{subfigure}[t]{.2\textwidth}
\includegraphics[width=.95\linewidth]{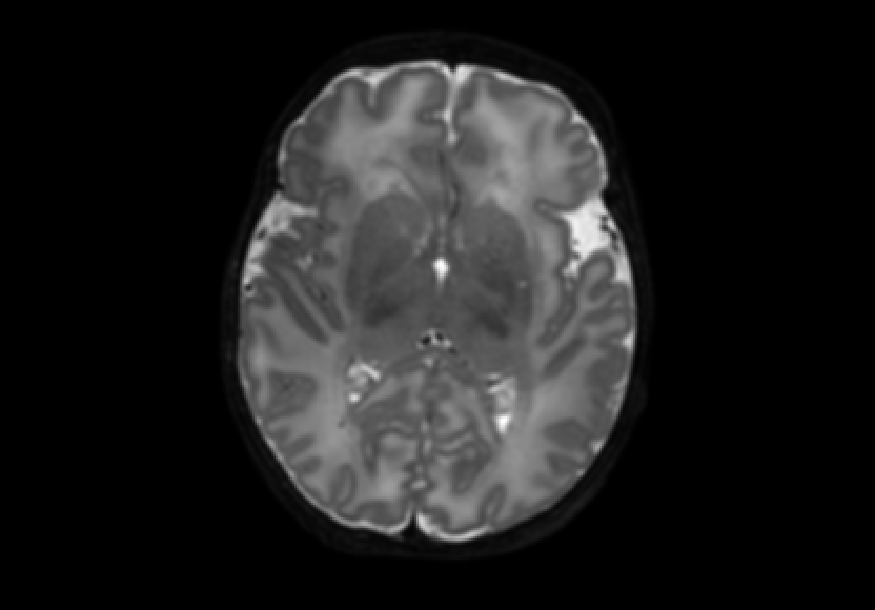}
\caption{$\sigma$=1}
\end{subfigure}%
\begin{subfigure}[t]{.2\textwidth}
\includegraphics[width=.95\linewidth]{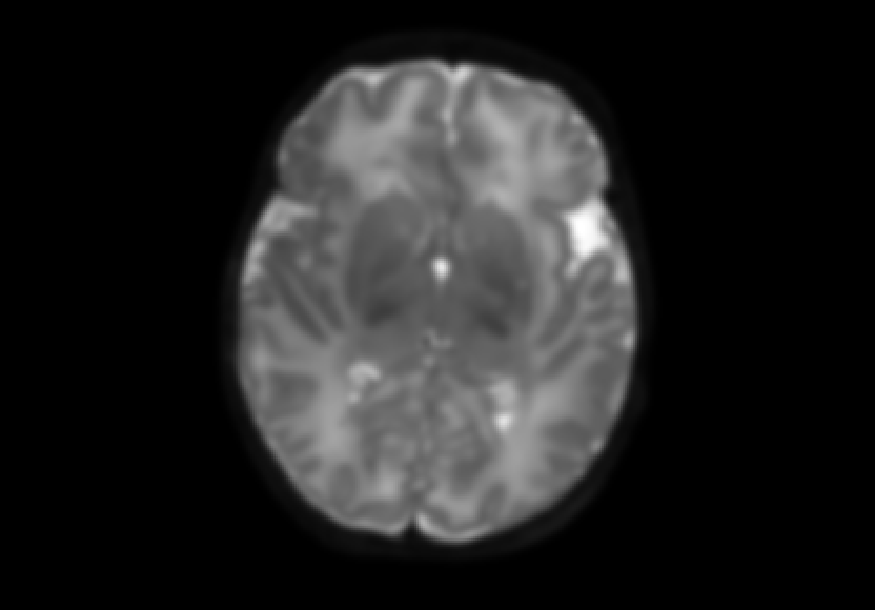}
\caption{$\sigma$=2}
\end{subfigure}%
\begin{subfigure}[t]{.2\textwidth}
\includegraphics[width=.95\linewidth]{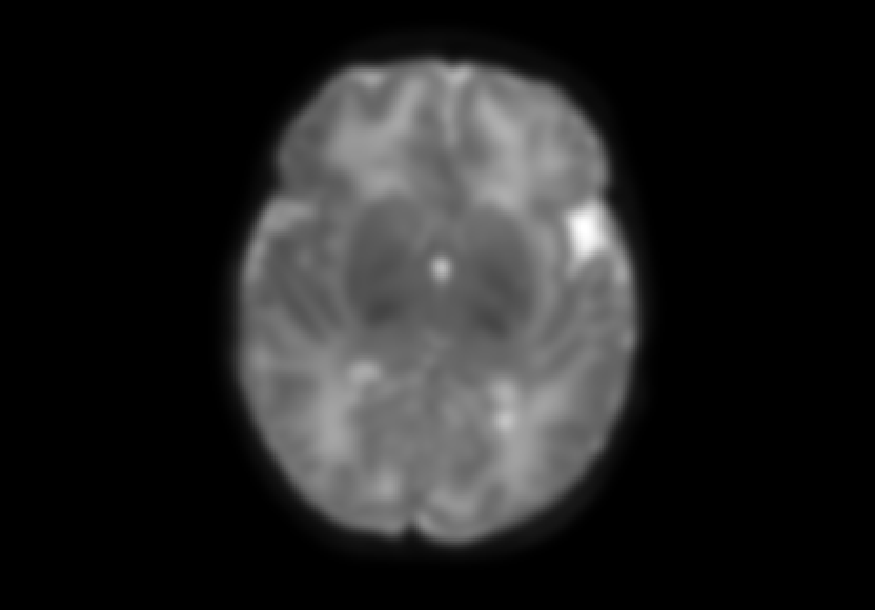}
\caption{$\sigma$=3}
\end{subfigure}%
\begin{subfigure}[t]{.2\textwidth}
\includegraphics[width=.95\linewidth]{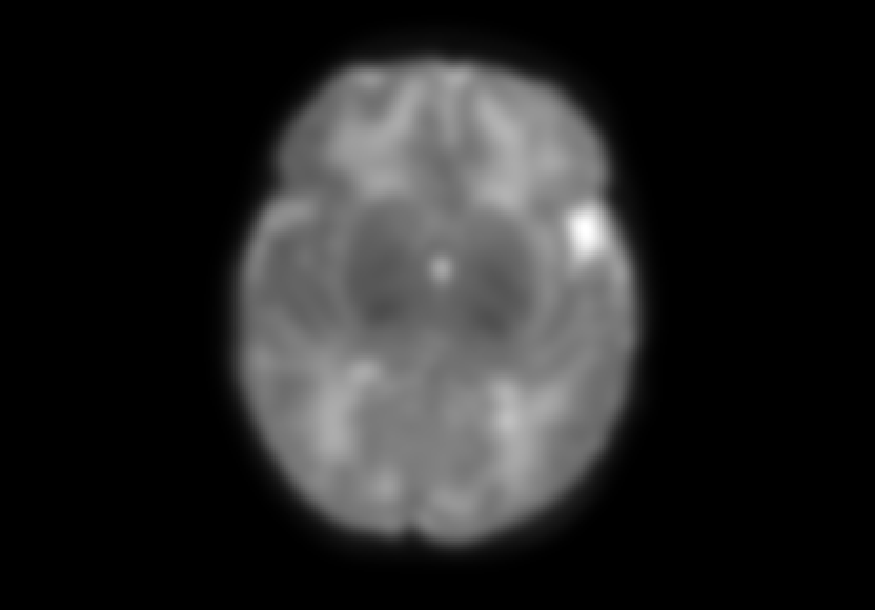}
\caption{$\sigma$=4}
\end{subfigure}%
\begin{subfigure}[t]{.2\textwidth}
\includegraphics[width=.95\linewidth]{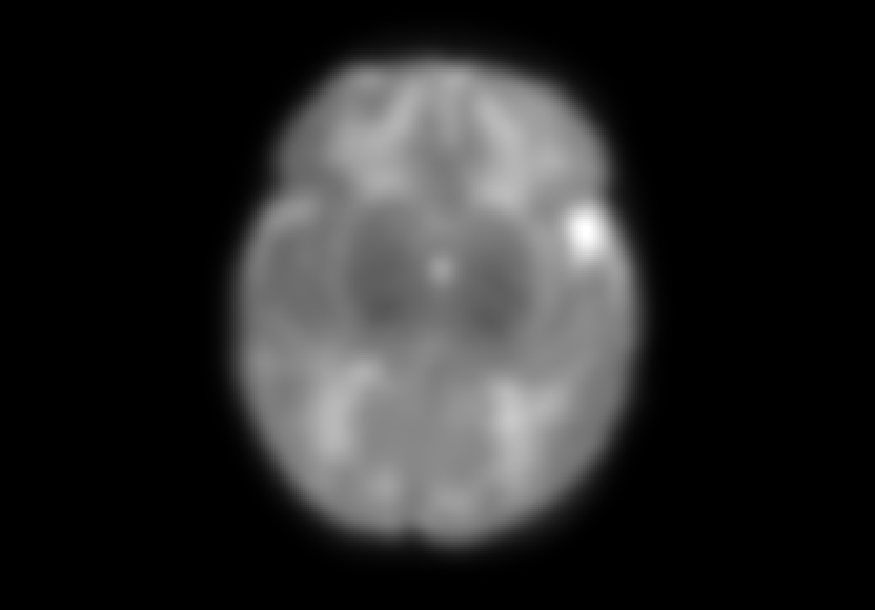}
\caption{$\sigma$=5}
\end{subfigure}
\caption{Axial slices of Gaussian blurred brain images with different values of $\sigma$.}\label{fig:dHCP_gaus_dataset}
\end{figure}

\textit{Blur} is a rather unique family of MRI artefacts because it can be introduced into a scan after the acquisition stage, such as during post-processing or in the manifestation of pathological conditions \cite{osadebey2018blind}; simulating blur textural artefacts was therefore important and relevant for this study. We used the well-established Gaussian blur, which can be produced by continuously applying a Gaussian filter to the image. Different degrees of blurring can be obtained by altering the $\sigma$ parameter, where higher values of $\sigma$ give blurrier transformations. We employed Gaussian filters from the library \texttt{scikit-image}~\cite{scikit-image} with $\sigma$ = \{1, 2, 3, 4, 5\} to give five different degrees of Gaussian blurred datasets named gaus01-05, respectively. Figure~\ref{fig:dHCP_gaus_dataset} shows examples of noisy images. 

\begin{figure}[h]
\centering
\begin{subfigure}[t]{.2\textwidth}
\includegraphics[width=.95\linewidth]{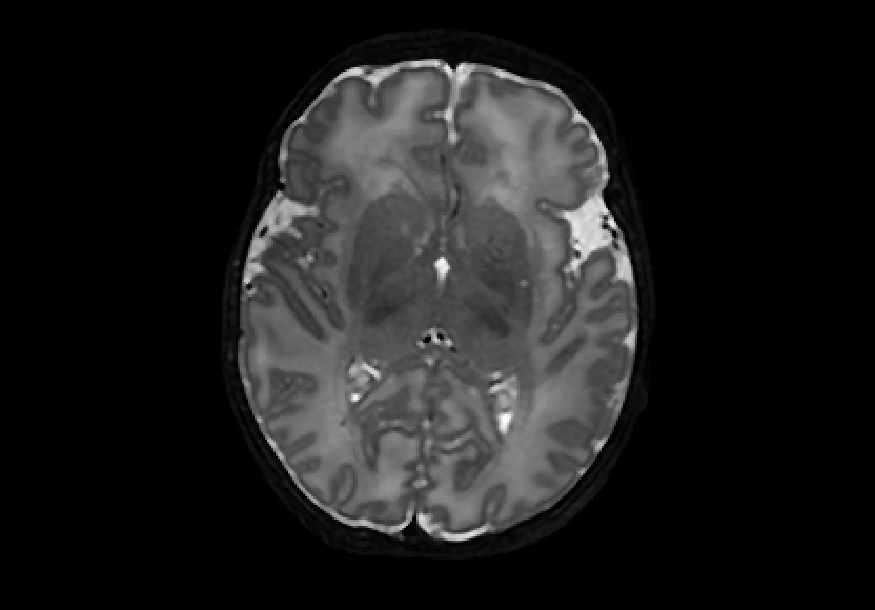}
\caption{\texttt{size} = 2}
\end{subfigure}%
\begin{subfigure}[t]{.2\textwidth}
\includegraphics[width=.95\linewidth]{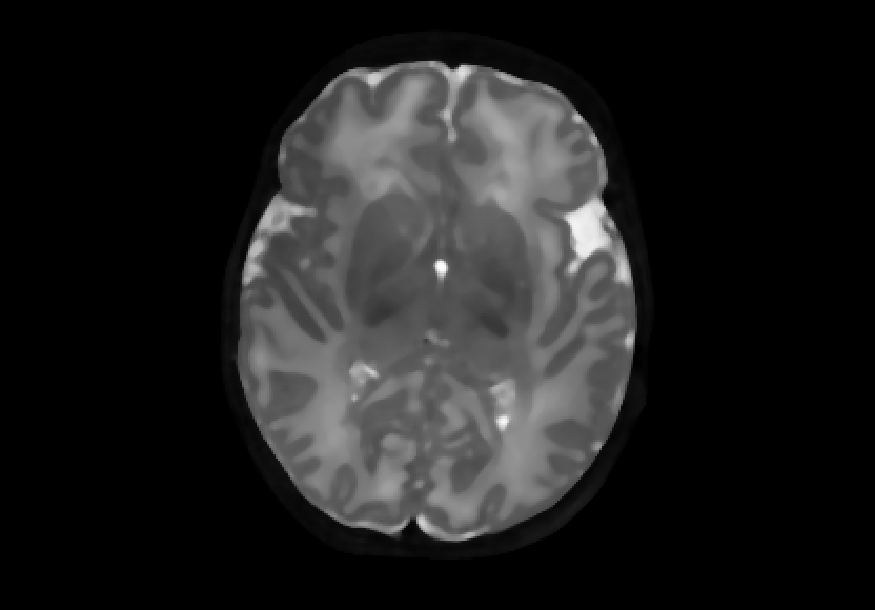}
\caption{\texttt{size} = 5}
\end{subfigure}%
\begin{subfigure}[t]{.2\textwidth}
\includegraphics[width=.95\linewidth]{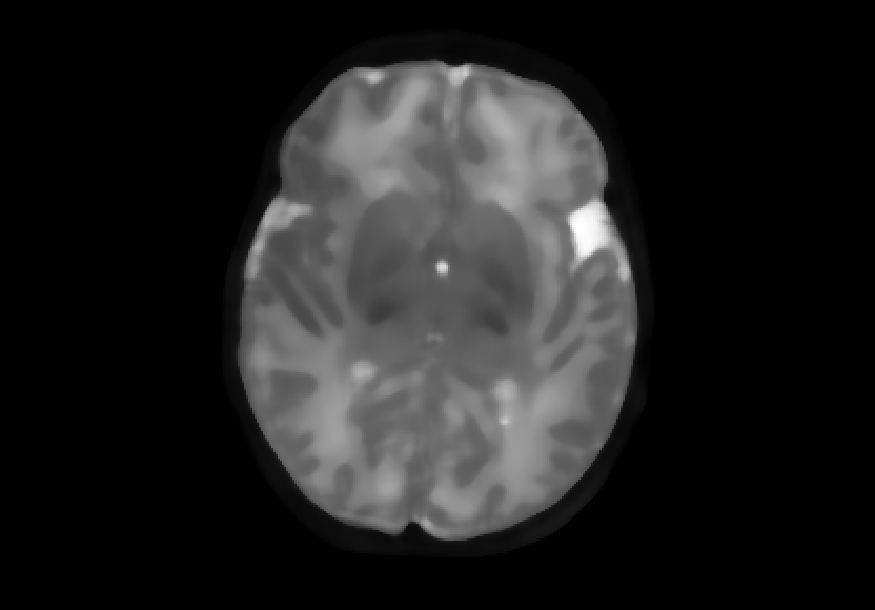}
\caption{\texttt{size} = 8}
\end{subfigure}
\caption{Axial slices of median smoothed images with different values of \texttt{size}.}
\label{fig:dHCP_median_dataset}
\end{figure}

Second, we simulated further blur manifestations using median filters, which replace the pixel value with the median of the neighbouring pixels. We utilised filters from \texttt{scipy}~\cite{scipy}, each with a parameter \texttt{size} that specifies the neighbourhood distance used to compute the median where higher values of \texttt{size} result in smoother filter output. Three different degrees of median filtered images with \texttt{size}= \{2, 5, 8\} were generated. Figure~\ref{fig:dHCP_median_dataset} shows examples of median-filtered images generated in this study.

\begin{figure}[h]
\centering
\begin{subfigure}[t]{.2\textwidth}
\includegraphics[width=.95\linewidth]{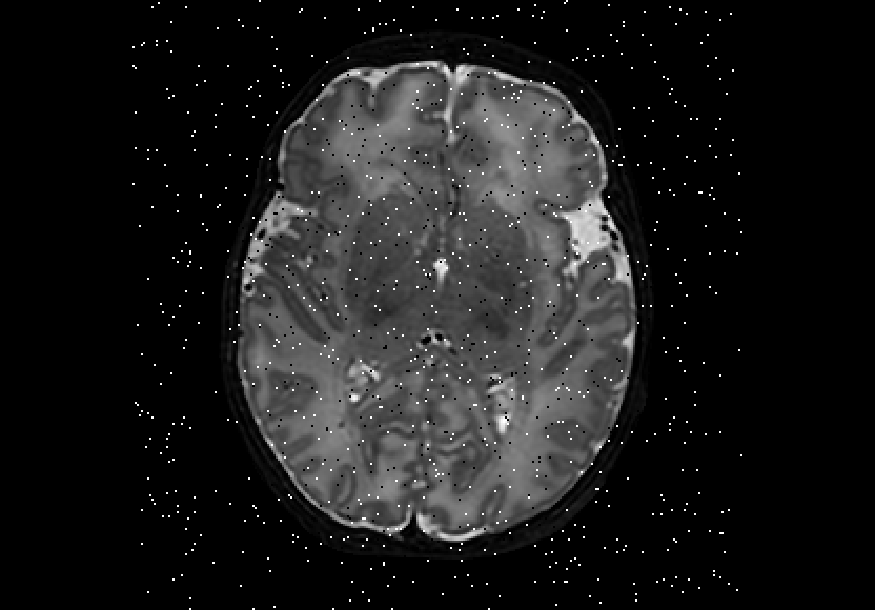}
\caption{\texttt{prob}=0.01}
\end{subfigure}%
\begin{subfigure}[t]{.2\textwidth}
\includegraphics[width=.95\linewidth]{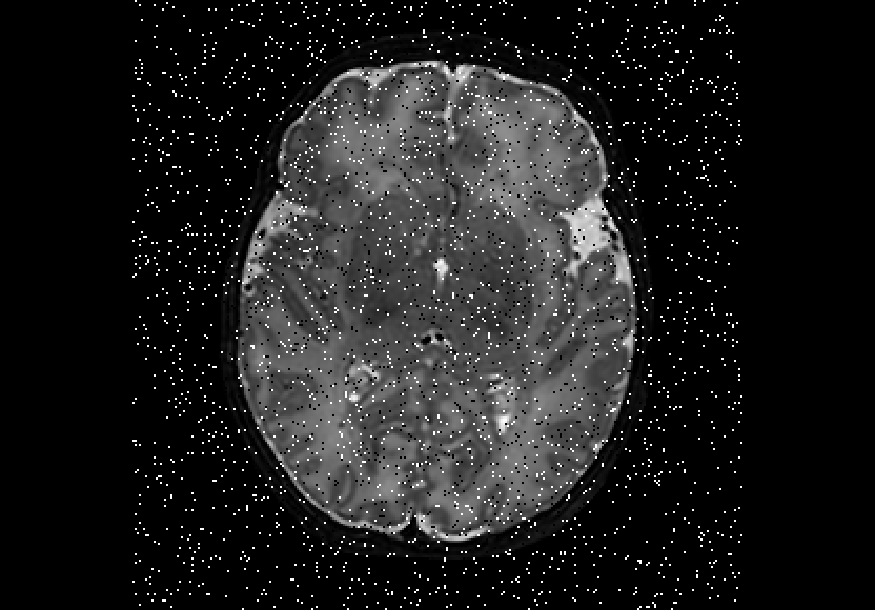}
\caption{\texttt{prob}=0.03}
\end{subfigure}%
\begin{subfigure}[t]{.2\textwidth}
\includegraphics[width=.95\linewidth]{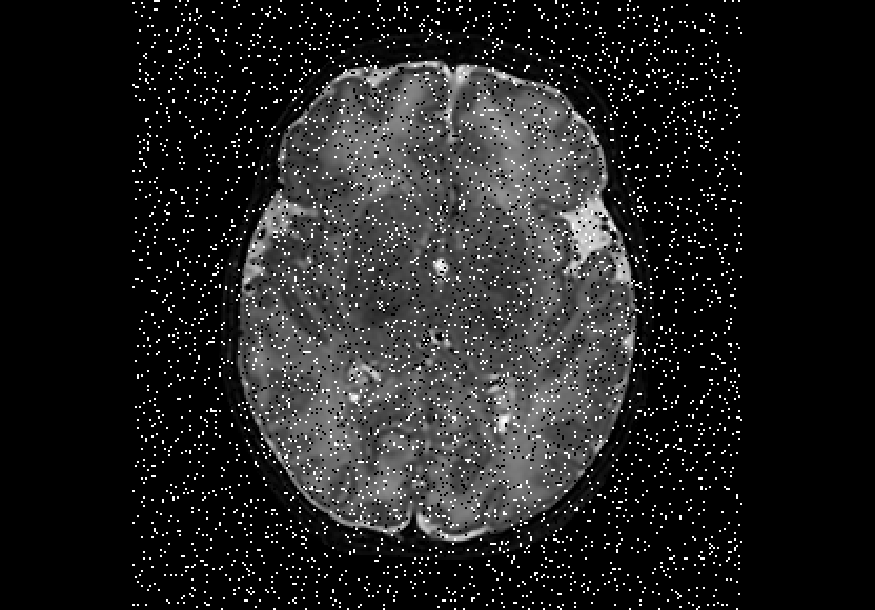}
\caption{\texttt{prob}=0.05}
\end{subfigure}%
\begin{subfigure}[t]{.2\textwidth}
\includegraphics[width=.95\linewidth]{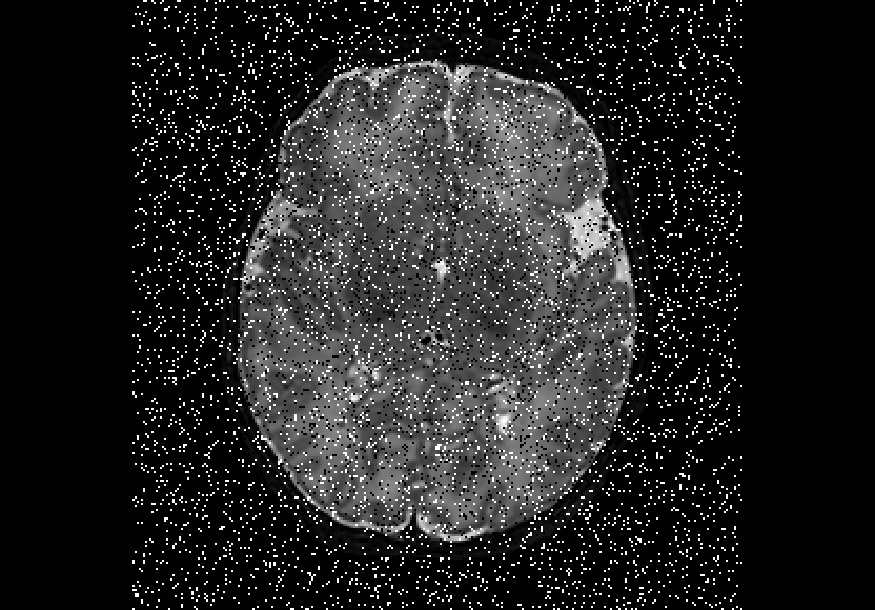}
\caption{\texttt{prob}=0.07}
\end{subfigure}

\begin{subfigure}[t]{.2\textwidth}
\includegraphics[width=.95\linewidth]{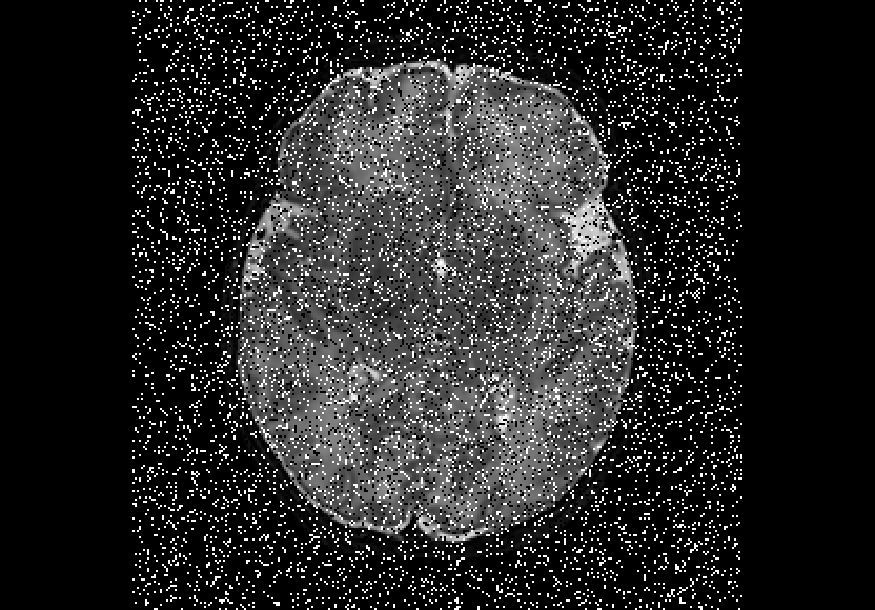}
\caption{\texttt{prob}=0.10}
\end{subfigure}%
\begin{subfigure}[t]{.2\textwidth}
\includegraphics[width=.95\linewidth]{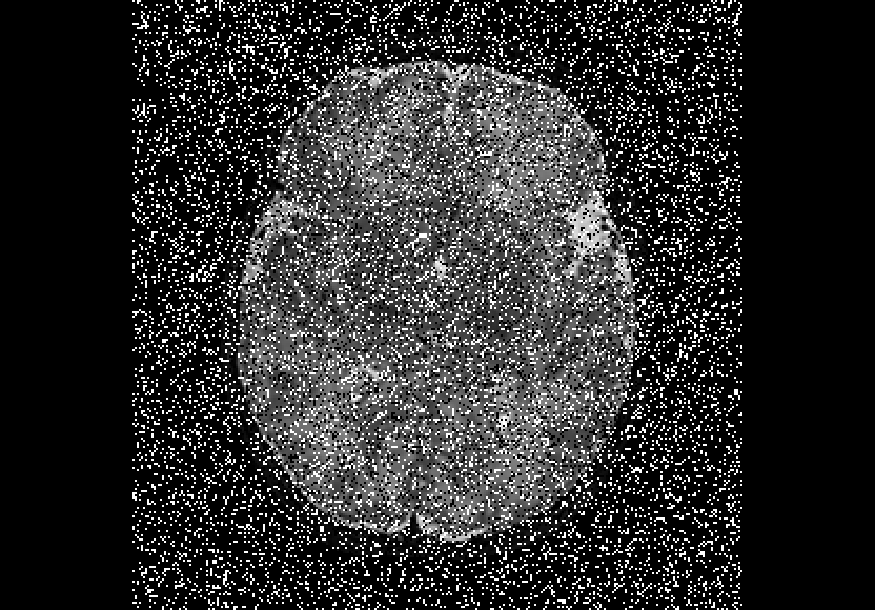}
\caption{\texttt{prob}=0.15}
\end{subfigure}%
\begin{subfigure}[t]{.2\textwidth}
\includegraphics[width=.95\linewidth]{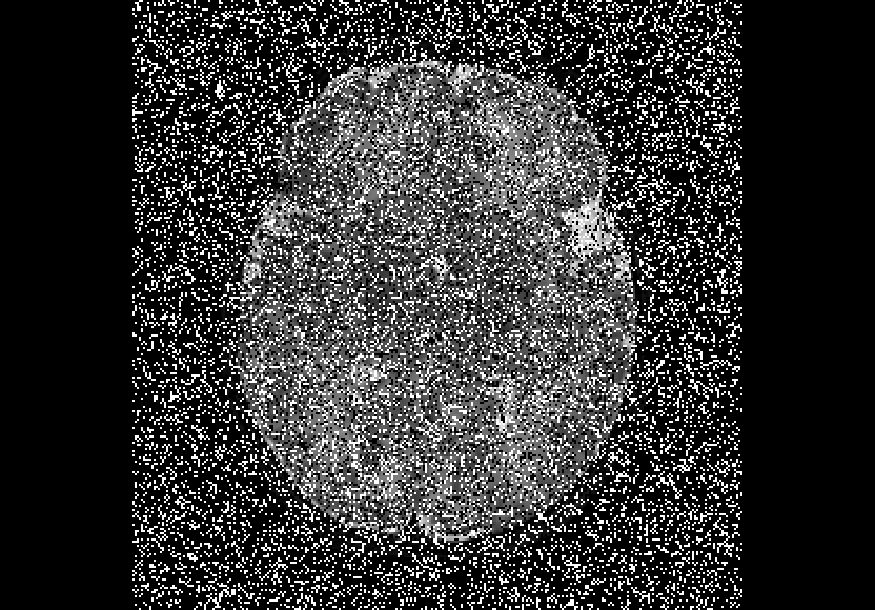}
\caption{\texttt{prob}=0.20}
\end{subfigure}
\caption{Axial slices of images injected with salt-and-pepper noise with different \texttt{prob} values.}
\label{fig:dHCP_snp_dataset}
\end{figure}

We also explored \textit{impulse noise} corruptions, such as those introduced by noisy communications channels, faulty memory locations, or damage in channel decoders~\cite{mousavi2017robust}. The impulse noise generating filter we used in this study is based on the salt-and-pepper (SNP) technique, which randomly generates black and white pixels on the image of interest. The function we used takes into account a parameter called \texttt{prob}, where 0 $\leq$ \texttt{prob} $\leq$  0.5. A random number is generated for each pixel; if it is less than \texttt{prob} then the function paints the pixel with black, if it is greater than \texttt{1-prob} then it paints the pixel with white, otherwise the pixel is left unchanged. In other words, the higher the value of \texttt{prob}, the noisier the output can become. We used different values for \texttt{prob}, in particular \texttt{prob} =\{0.01, 0.03, 0.05, 0.07, 0.10, 0.15, 0.20\}, to create seven different noisy datasets named snp\_prob. Examples of axial slices are shown in Figure~\ref{fig:dHCP_snp_dataset}.

It is important to note that when we carried out the experiments, which are discussed in Section \ref{experiments}, we reserved some of the generated noisy datasets solely for use at test time. This was done in order to simulate various degrees of domain shift encountered in routine clinical practice (e.g. changes in magnetic field strength, scanner hardware, acquisition protocols, hospital's digital communications infrastucture) that may not be experienced during training.

\subsection{CNN Architecture} \label{cnn}

We used DeepMedic, an open source 3D architecture with multiple convolutional pathways originally built for brain lesion segmentation tasks \cite{kamnitsas2017_deepmedic}. Our CNN had eight layers in both the normal pathway and the subsampled pathways with the kernel size of $3^3$. Two parallel subsampled pathways were used, giving a total of three pathways. We also employed residual connections between layers 3 and 4, layers 5 and 6, and layers 7 and 8. The number of feature maps used in each fully connected layer was 250. We ran the experiment for 100 epochs where each epoch comprised of 20 subepochs. In every subepoch, images from five cases were loaded and the training samples were extracted. We used a batch size of 10, in addition to a predefined learning rate scheduler which schedules every 8 epochs starting from the 24th epoch. Finally, we used RMSProp as an optimiser, and L1 and L2 regularisation with values $10^{-5}$ and $10^{-4}$.

\subsection{Experiments} \label{experiments}

Of the 70 scans provided, we randomly selected 50 scans for training,  10 for validation, and 10 for testing. This is a fairly complex tissue segmentation task because visual characteristics of the scans can greatly vary throughout brain development, e.g. as the brain matures, the occurrence of dark intensities present in white matter regions gradually increases. The Dice similarity coefficient (DSC) metric was used for model evaluation.

Since the motivation of the experiments was to find out whether training models under a certain textural noise setting could potentially help with domain shift, some of the noisy datasets created above were intentionally not used during model training. The CNN was trained on 11 different combinations of the following datasets: t2original, gaus01, gaus03, gaus04, snp01, snp05, snp10 and median05, as summarised in Table~\ref{table:medical_models}. 

Each of the 11 models was then evaluated on 16 held-out test sets. The total number of experiments carried out was therefore 176. Note that the unseen noise types/levels reserved purely for testing were gaus02, gaus05, snp03, snp07, snp15, snp20, median02, and median08. The use of unseen noise levels at test time is important to simulate realistic domain shifts introduced in healthcare settings, e.g. changes in magnetic field strength or scanner hardware. Our goal was to find out whether training on certain combinations of textural noise can lead to \textit{textural invariant} models, and hence lead to robustness on new, previously unseen noise. 

\begin{table}[h]
\centering
\caption{Model names and the corresponding combinations of training data.}\label{table:medical_models}
\begin{tabular}{|l|l|}
\hline
Model Name                  & Training Data Combination                          \\
\hline
M\_T2ORIGINAL              & t2original (baseline, noise free)                                    \\
M\_GAUS01                & gaus01 ($\sigma$= 1)                                     \\
M\_GAUS03                & gaus03 ($\sigma$= 3)                                     \\
M\_GAUS04                & gaus04  ($\sigma$= 4)                                    \\
M\_GAUS010304              & gaus01, gaus03, gaus04  ($\sigma$= 1, 3, 4)                      \\
M\_SNP01                & snp01 (\texttt{prob} = 0.01)                                    \\
M\_SNP05                & snp05 (\texttt{prob} = 0.05)                                   \\
M\_SNP10                & snp10 (\texttt{prob} = 0.10)                                  \\
M\_SNP010510            & snp01, snp05, snp10 (\texttt{prob} = 0.01, 0.05, 0.10)\\
M\_MEDIAN05              & median05 (\texttt{size} = 5)                      \\
M\_GAUS010304\_SNP010510   & gaus01, gaus03, gaus04 ($\sigma$= 1, 3, 4), \\ 
&  snp01, snp05, snp10 (\texttt{prob} = 0.01, 0.05, 0.10)   \\

\hline
\end{tabular}
\end{table}

\section{Results and Discussion} \label{results}

All 11 models were first tested on held out data that represented the settings they were trained on. The models were then tested on held out data that represented settings which are completely new and are assumed to simulate domain shift. For succinctness, we present in detail 3 of the 11 models, hence showing the most insightful results in the tables below. The reported DSC values corresponded to the following tissue classes: \textit{1)} zero-intensity background, \textit{2)} CSF, \textit{3)} cGM, \textit{4)} WM, \textit{5)} background bordering brain tissues, \textit{6)} ventricles, \textit{7)} cerebellum, \textit{8)} dGM, \textit{9)} brainstem, and \textit{10)} hippocampus. 

We observed that the model trained on noise-free scans achieved highest overall robustness on data similar to the data it was trained on and where textural transformation was only lightly applied; a rather expected finding (see Table \ref{table:model_t2norm_dsc}). For instance, the noise-free model was able to achieve a DSC of over 86\% for all 10 classes on gaus01, but its performance immensely dropped to 0-1\% with 4 of those classes on gaus05. This shows that the baseline model trained using a conventional approach does not immediately generalise to unseen noise categories and levels, highlighting the problem of domain shift often experienced with a wide range of realistic scenarios, e.g. variations in scanner manufacturer, magnetic field strengths, or acquisition protocols. 

\begin{table}
\centering
\caption{DSCs achieved by the baseline, noise-free model (M\_T2ORIGINAL) on the 16 versions of the held-out test set. Data highlighted in bold correspond to settings where the model achieves over 90\% DSC for \textit{all} classes, demonstrating particularly high levels of robustness. The model achieved highest overall robustness on test sets that are noise-free and where textural transformation was only lightly applied, reflecting the common problem of domain shift in imaging.}\label{table:model_t2norm_dsc}
\begin{tabu}{|l|l|l|l|l|l|l|l|l|l|l|}
\hline
Data/Class   & 1 & 2 & 3 & 4 & 5 & 6 & 7 & 8 & 9 & 10 \\
\hline
\textbf{t2original}  & 99.35 & 95.99 & 96.64 & 97.57 & 91.96 & 95.39 & 97.78 & 97.06 & 97.12 & 92.98 \\
gaus01   & 98.87 & 90.21 & 91.97 & 94.30 & 86.43 & 90.79 & 95.97 & 95.28 & 95.67 & 89.90 \\
gaus02   & 96.69 & 54.18 & 33.40 & 65.59 & 70.94 & 62.57 & 59.77 & 56.65 & 88.93 & 66.08 \\
gaus03   & 94.98 & 15.67 & 03.84 & 51.65 & 60.83 & 25.27 & 44.42 & 42.43 & 69.00 & 13.47 \\
gaus04   & 94.71 & 00.78 & 01.75 & 53.41 & 54.99 & 00.25 & 33.44 & 45.49 & 16.04 & 00.00 \\
gaus05   & 93.98 & 00.19 & 01.61 & 47.00 & 45.44 & 00.08 & 35.81 & 36.73 & 00.33 & 00.05 \\
snp01   & 99.22 & 87.14 & 85.02 & 84.45 & 89.62 & 88.95 & 94.02 & 86.11 & 92.51 & 83.83 \\
snp03   & 93.96 & 45.62 & 48.08 & 14.73 & 63.21 & 19.28 & 16.85 & 05.85 & 47.76 & 10.85 \\
snp05   & 55.33 & 26.53 & 06.63 & 01.60 & 23.90 & 00.32 & 00.01 & 00.33 & 01.79 & 00.17 \\
snp07   & 16.99 & 21.49 & 00.03 & 00.08 & 12.38 & 00.10 & 00.00 & 00.02 & 00.01 & 00.01 \\
snp10   & 03.03 & 07.80 & 00.00 & 00.00 & 04.05 & 00.00 & 00.00 & 00.00 & 00.00 & 00.00 \\
snp15   & 00.15 & 00.53 & 00.00 & 00.00 & 00.24 & 00.00 & 00.00 & 00.00 & 00.00 & 00.00 \\
snp20   & 00.02 & 00.12 & 00.00 & 00.00 & 00.01 & 00.00 & 00.00 & 00.00 & 00.00 & 00.00 \\
median02 & 98.91 & 87.54 & 88.86 & 92.58 & 81.99 & 88.95 & 96.22 & 95.22 & 94.11 & 88.33 \\
median05 & 99.18 & 85.15 & 86.57 & 91.43 & 88.50 & 86.17 & 89.59 & 91.57 & 94.81 & 86.82 \\
median08 & 98.38 & 61.75 & 46.73 & 73.58 & 76.40 & 59.42 & 57.81 & 63.40 & 87.58 & 48.14 \\
\hline
\end{tabu}
\end{table}

A similar pattern emerged when we trained models on images smoothed out by the Gaussian filter only, we observed that they generalised on Gaussian filtered images that were generated using values of $\sigma$ close to those used on training data. For example, a model which was trained only on Gaussian noise  $\sigma$=3 
showed the highest overall DSC values on datasets gaus03 and gaus04, but not on other variations of the Gaussian images.
Noteworthy, we also observed that when a model was trained on a Gaussian dataset created with a very large value for $\sigma$, it did not segment the noise-free data correctly. We believe this is because the use of such blur filters alone may have completely distorted useful anatomical features within the scans. Interestingly, however, when a model was trained on multiple Gaussian datasets generated with a variety of $\sigma$ values, it generalised well on the noise-free data as well as on Gaussian filtered images, even for the cases where the $\sigma$ values used were different from that used in training. Moreover, segmentation performance on the median filtered data also increased. Nevertheless, the models did not result in high DSCs for the Gaussian datasets with the highest $\sigma$ value or the salt-and-pepper noise. 
This behaviour was consistent with the performance of models trained solely on median filtered images, 
which showed low levels of robustness on heavier filtered median image and Gaussian filtered images, as well as a lack of generalisation when tested on salt-and-pepper filtered images.

Interestingly, and unlike the Gaussian image trained models, models trained on salt-and-pepper noise showed high levels of robustness on all the lighter noise levels, some heavier noise levels, as well as completely different noise categories. For instance, model M\_SNP01 (fairly low levels of salt-and-pepper noise, \texttt{prob} = 0.01) showed high levels of robustness on data with heavier salt-and-pepper noise, on the original dHCP images, as well as on the lightly texturised Gaussian and median images. This was also consistent with the robustness observed with M\_SNP10 (trained on very high levels of salt-and-pepper noise, \texttt{prob}=0.10) which is detailed in Table \ref{table:model_snp10_dsc}. 

Our findings therefore suggest that models trained on impulse noise corruptions, namely salt-and-pepper textured images, can generalise well on images similarly corrupted by salt-and-pepper noise as well as on blur smoothed images. This leads us to believe that the model developed invariance to low frequency textural patterns and hence used relatively useful information about the global anatomical structure of the brain when carrying out tissue segmentation. Models trained on blur smoothed images, however, only generalised to images smoothed to similar degrees but not on images smoothed using other techniques.

\begin{table}[h]
\centering
\caption{DSCs achieved by the model trained on high levels of salt-and-pepper noise (M\_SNP10, \texttt{prob} = 0.01) on the 16 versions of the held-out test set. Data highlighted in bold correspond to settings where the model achieves over 90\% DSC for \textit{all} classes, demonstrating particularly high levels of robustness. A key observation is that this model demonstrated high levels of robustness on all 10 classes for noise-free data (t2original), data injected with lower levels of salt-and-pepper noise (e.g. snp01), as well as higher levels of salt-and-pepper noise (e.g. snp15). }\label{table:model_snp10_dsc}
\begin{tabu}{|l|l|l|l|l|l|l|l|l|l|l|}
\hline
Data/Class   & 1 & 2 & 3 & 4 & 5 & 6 & 7 & 8 & 9 & 10 \\
\hline
\textbf{t2original}  & 99.30 & 95.17 & 95.84 & 96.80 & 91.15 & 94.20 & 97.25 & 96.39 & 96.38 & 91.41 \\
gaus01   & 98.86 & 90.37 & 91.58 & 93.74 & 86.45 & 90.80 & 95.66 & 94.66 & 94.77 & 88.77 \\
gaus02   & 97.32 & 50.32 & 55.60 & 69.55 & 71.64 & 73.49 & 87.67 & 80.20 & 89.31 & 81.33 \\
gaus03   & 96.07 & 20.73 & 16.64 & 59.98 & 58.44 & 52.44 & 83.01 & 69.76 & 83.82 & 71.18 \\
gaus04   & 95.55 & 08.15 & 07.77 & 57.44 & 48.09 & 25.70 & 77.99 & 62.70 & 71.66 & 57.42 \\
gaus05   & 95.48 & 03.40 & 06.46 & 56.26 & 37.80 & 12.07 & 70.84 & 56.04 & 47.10 & 41.67 \\ 
\textbf{snp01}   & 99.31 & 95.34 & 95.97 & 96.94 & 91.33 & 94.61 & 97.34 & 96.54 & 96.59 & 91.73 \\ 
\textbf{snp03}   & 99.31 & 95.59 & 96.18 & 97.13 & 91.56 & 94.98 & 97.45 & 96.74 & 96.82 & 92.26 \\ 
\textbf{snp05}   & 99.32 & 95.75 & 96.31 & 97.27 & 91.65 & 95.11 & 97.51 & 96.81 & 96.85 & 92.50 \\ 
\textbf{snp07}   & 99.32 & 95.83 & 96.39 & 97.36 & 91.69 & 95.00 & 97.51 & 96.84 & 96.88 & 92.71 \\ 
\textbf{snp10}   & 99.33 & 95.86 & 96.39 & 97.39 & 91.67 & 95.14 & 97.56 & 96.84 & 96.93 & 92.72 \\ 
\textbf{snp15}   & 99.32 & 95.51 & 95.92 & 97.11 & 91.40 & 94.51 & 97.39 & 96.66 & 96.73 & 92.08 \\
snp20   & 99.04 & 94.08 & 94.37 & 96.09 & 89.12 & 93.57 & 96.53 & 95.59 & 95.99 & 89.69 \\
median02 & 98.87 & 87.03 & 89.04 & 92.68 & 82.00 & 87.89 & 95.69 & 94.81 & 93.65 & 87.74 \\
median05 & 99.17 & 84.75 & 87.21 & 91.36 & 88.14 & 87.05 & 93.27 & 92.65 & 93.56 & 86.79 \\
median08 & 98.63 & 62.46 & 59.60 & 76.26 & 78.12 & 68.34 & 84.49 & 83.03 & 87.76 & 78.60 \\
\hline
\end{tabu}
\end{table}

\begin{figure}[h]
\centering
\begin{subfigure}[t]{.25\textwidth}
\centering
\includegraphics[width=.95\linewidth]{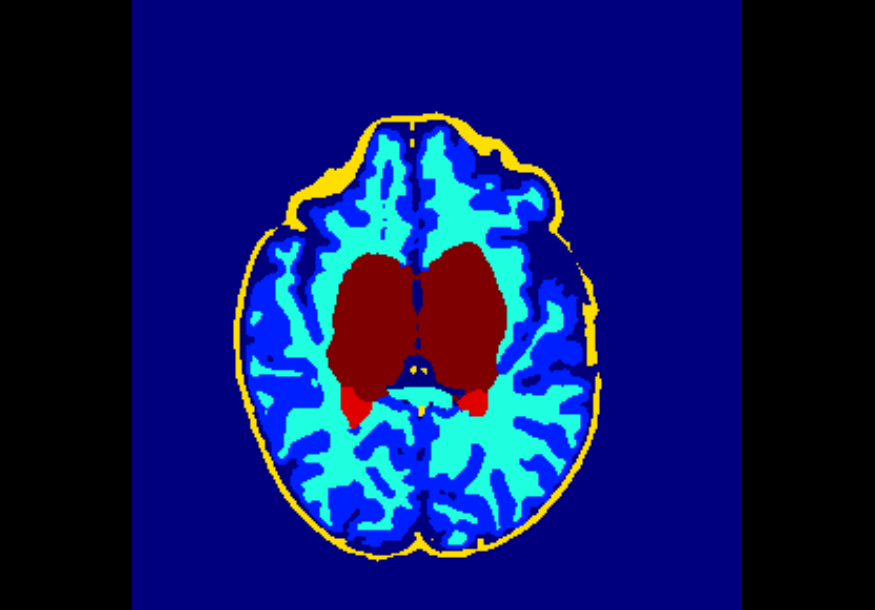}
\caption{Ground-truth}
\end{subfigure}%
\begin{subfigure}[t]{.25\textwidth}
\centering
\includegraphics[width=.95\linewidth]{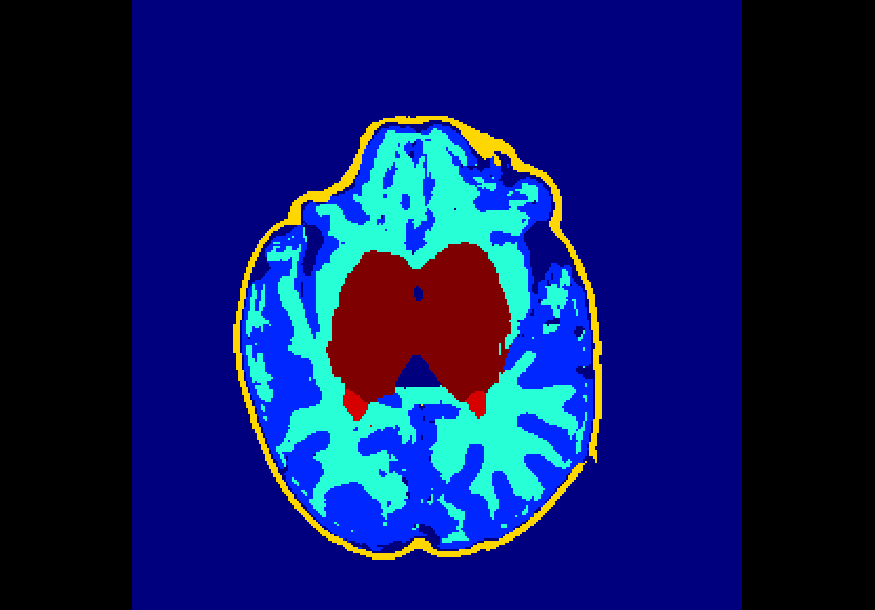}
\caption{Output:median08 }
\end{subfigure}%
\begin{subfigure}[t]{.25\textwidth}
\centering
\includegraphics[width=.95\linewidth]{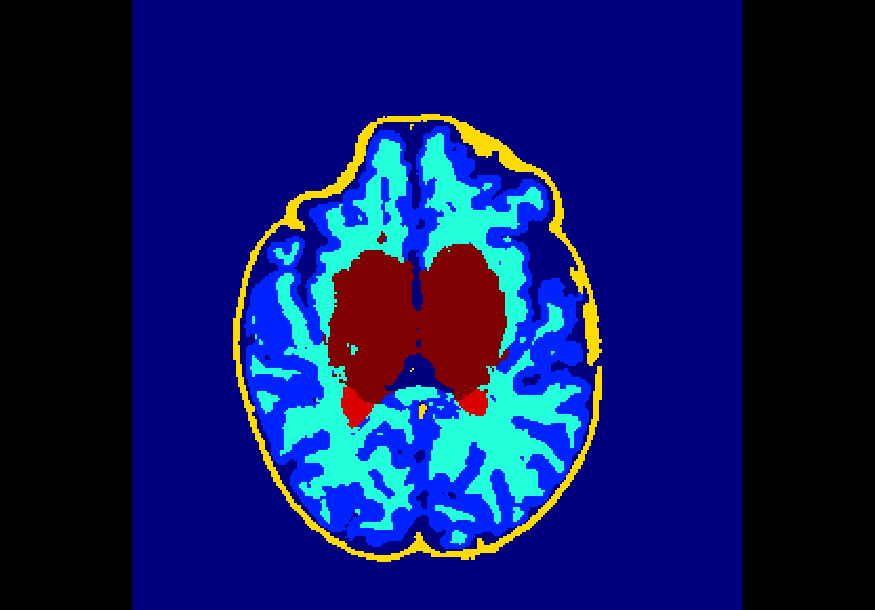}
\caption{Output: snp20 }
\end{subfigure}%
\begin{subfigure}[t]{.25\textwidth}
\centering
\includegraphics[width=.95\linewidth]{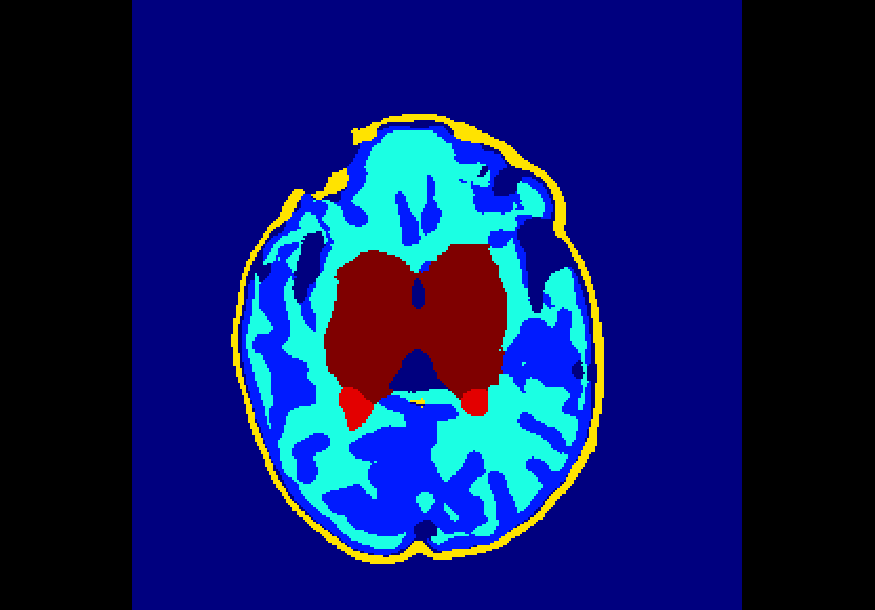}
\caption{Output: gaus05 }
\end{subfigure}
\caption{Example performance of the model trained on several levels of Gaussian and salt-and-pepper noise
on previously unseen textural noise levels.}\label{fig:gaus134_snp010510_prediction}
\end{figure}

Finally, Table~\ref{table:model_gaus134_snp010510_dsc} shows that the model trained on data injected with different degrees of both Gaussian and salt-and-pepper noise 
(gaus01, gaus03, gaus04 with $\sigma$= 1, 3, 4 ; snp01, snp05, snp10 with \texttt{prob} = 0.01, 0.05, 0.10) achieves the best overall robustness across the 16 versions of the test set, where even the heaviest filtered images had very few segmentation inaccuracies. For instance, this model demonstrated high levels of robustness on all 10 classes for noise-free data, on data injected with a previously unseen degree of salt-and-pepper noise (e.g. DSCs of 88\%-99\% on snp20), as well as on data with a previously unseen degree of Gaussian noise (e.g. DSCs of 72\%-98\% on gaus05). To illustrate, Figure~\ref{fig:gaus134_snp010510_prediction} shows examples of the predicted segmentation of the model on the heaviest transformed images of the three filter categories used in these experiments. From the figure it is clear that the model achieved excellent mapping of the brain tissue regions. This shows a tremendous improvement from the baseline model which was trained using a conventional, noise-free approach and thus failed severely on the heavily corrupted images (e.g. DSCs of 0\% for all 10 classes on snp20, see Table \ref{table:model_t2norm_dsc}).

\newpage
\begin{table}[h]
\centering
\caption{DSCs achieved by the model trained on several levels of Gaussian ($\sigma$= 1, 3, 4) and salt-and-pepper noise (\texttt{prob} = 0.01, 0.05, 0.10) on the 16 versions of the held-out test set. Data highlighted in bold correspond to settings where the model achieves over 0.9 DSC for \textit{all} classes, demonstrating particularly high levels of robustness. This model exhibited the best overall robustness across all levels and types of noise tested, as well as on all 10 classes studied.}\label{table:model_gaus134_snp010510_dsc}
\begin{tabu}{|l|l|l|l|l|l|l|l|l|l|l|}
\hline
Data/Class   & 1 & 2 & 3 & 4 & 5 & 6 & 7 & 8 & 9 & 10 \\
\hline
\textbf{t2original}  & 99.28 & 94.62 & 95.21 & 96.45 & 90.83 & 94.24 & 97.18 & 96.34 & 96.14 & 92.11 \\ 
\textbf{gaus01}   & 99.27 & 94.52 & 95.00 & 96.34 & 90.64 & 94.47 & 97.24 & 96.36 & 96.19 & 92.06 \\
gaus02   & 99.18 & 89.87 & 90.89 & 92.64 & 88.05 & 91.37 & 96.21 & 95.68 & 95.36 & 90.83 \\
gaus03   & 99.10 & 89.26 & 89.68 & 92.25 & 86.41 & 91.74 & 96.15 & 95.49 & 95.07 & 90.00 \\
gaus04   & 98.97 & 85.77 & 85.94 & 89.14 & 83.92 & 90.28 & 95.52 & 94.88 & 94.41 & 88.29 \\
gaus05   & 98.56 & 72.98 & 73.12 & 79.46 & 75.91 & 82.91 & 92.77 & 92.60 & 91.74 & 83.86 \\ 
\textbf{snp01}   & 99.28 & 94.87 & 95.42 & 96.62 & 90.92 & 94.43 & 97.26 & 96.38 & 96.19 & 92.09 \\ 
\textbf{snp03}   & 99.28 & 95.06 & 95.56 & 96.73 & 91.00 & 94.56 & 97.30 & 96.44 & 96.20 & 92.01 \\ 
\textbf{snp05}   & 99.28 & 95.10 & 95.57 & 96.71 & 90.99 & 94.50 & 97.32 & 96.39 & 96.22 & 91.78 \\ 
\textbf{snp07}   & 99.27 & 95.11 & 95.54 & 96.66 & 90.92 & 94.43 & 97.21 & 96.36 & 96.27 & 91.68 \\ 
\textbf{snp10}   & 99.26 & 95.06 & 95.42 & 96.55 & 90.78 & 94.41 & 97.19 & 96.21 & 96.11 & 91.64 \\ 
\textbf{snp15}   & 99.23 & 94.56 & 94.63 & 95.86 & 90.25 & 92.84 & 96.80 & 95.48 & 95.49 & 90.03 \\
snp20   & 99.07 & 92.99 & 91.95 & 93.74 & 88.33 & 88.39 & 94.86 & 92.96 & 94.40 & 86.02 \\
median02 & 98.90 & 87.01 & 88.24 & 92.08 & 81.71 & 88.41 & 95.92 & 94.57 & 93.42 & 87.88 \\
median05 & 99.27 & 88.30 & 91.02 & 94.30 & 89.18 & 88.50 & 96.10 & 95.36 & 95.04 & 90.19 \\
median08 & 98.86 & 72.79 & 77.41 & 84.45 & 79.29 & 74.46 & 92.41 & 92.62 & 91.24 & 84.39 \\
\hline
\end{tabu}
\end{table}

\section{Conclusion} \label{sec:conclusion}

The medical domain contains all the factors that make machine learning algorithms challenging to translate to routine practice. Predictive models would have to adapt to unexpected circumstances when solving perceptual tasks or planning treatment strategies, due to the infinite variability in human nature and healthcare systems. In medical image segmentation, a major bottleneck that is often encountered when developing machine learning models is the issue of robustness under domain shift, where changes in hospital infrastructure, acquisition hardware, or image resolution cause seemingly accurate models to immensely fail at test time. 

In this study, we hypothesised that addressing the textural bias phenomenon in medical imaging can lead to CNNs that are \textit{texture invariant} and hence more resilient to changes in data distribution. Specifically, our motivation was  to  find  out  whether  training deep segmentation models  under the right textural noise settings could help improve model robustness. In this regard, we carried out an extensive empirical study consisting of 176 experiments for a complex brain tissue segmentation task, exploring the effect of training with several permutations of three main types of noise: Gaussian blur, median filter blur, and impulse salt-and-pepper noise. A key finding of our work is that training a deep segmentation model on neuroimaging data injected with certain combinations of textural noise can indeed improve model robustness on new, previously unseen noise levels. We believe that this is due to the models being incentified to learn anatomical and tissue-specific features, as opposed to low-frequency textural patterns that may be brittle and domain specific (e.g. inherent to a given scanner or acquisition protocol). 

In terms of extending the investigation further, the next natural step would be to conduct a similar experiment on even more permutations of smoothing, salt-and-pepper generating, or even contrast changing filters to see if training models in these settings can yield even better segmentation results. Additionally, an investigation on whether this hypothesis is valid for non-simulated domain shift of neuroimaging data - ideally using data acquired across different sites - ought to be carried out. If findings are consistent with this study, generalisation to further modalities would be important to explore.

\section{Acknowledgments}
The research leading to these results has received funding from the European Research Council under the European Union’s Seventh Framework Programme (FP/2007-2013)/ERC Grant Agreement no. 319456. We are grateful to the families who generously supported his trial. 
%
%

\bibliographystyle{splncs04}
\bibliography{bib}

\begin{thebibliography}{10}
\providecommand{\url}[1]{\texttt{#1}}
\providecommand{\urlprefix}{URL }
\providecommand{\doi}[1]{https://doi.org/#1}

\bibitem{bastiani2018}
Bastiani, M., Andersson, J.L.R., Cordero{-}Grande, L., Murgasova, M., Hutter,
  J., Price, A.N., Makropoulos, A., Fitzgibbon, S.P., Hughes, E.J., Rueckert,
  D., Victor, S., Rutherford, M.A., Edwards, A.D., Smith, S.M., Tournier, J.,
  Hajnal, J.V., Jbabdi, S., Sotiropoulos, S.N.: Automated processing pipeline
  for neonatal diffusion {MRI} in the developing human connectome project.
  NeuroImage  \textbf{185},  750--763 (2019),
  \url{https://doi.org/10.1016/j.neuroimage.2018.05.064}

\bibitem{deng2009imagenet}
Deng, J., Dong, W., Socher, R., Li, L.J., Li, K., Fei-Fei, L.: Imagenet: A
  large-scale hierarchical image database. In: 2009 IEEE conference on computer
  vision and pattern recognition. pp. 248--255. Ieee (2009)

\bibitem{geirhos2018}
Geirhos, R., Rubisch, P., Michaelis, C., Bethge, M., Wichmann, F.A., Brendel,
  W.: Imagenet-trained cnns are biased towards texture; increasing shape bias
  improves accuracy and robustness. In: 7th International Conference on
  Learning Representations, {ICLR}. OpenReview.net (2019),
  \url{https://openreview.net/forum?id=Bygh9j09KX}

\bibitem{kamnitsas2017}
Kamnitsas, K., Baumgartner, C.F., Ledig, C., Newcombe, V.F.J., Simpson, J.P.,
  Kane, A.D., Menon, D.K., Nori, A.V., Criminisi, A., Rueckert, D., Glocker,
  B.: Unsupervised domain adaptation in brain lesion segmentation with
  adversarial networks. In: Niethammer, M., Styner, M., Aylward, S.R., Zhu, H.,
  Oguz, I., Yap, P., Shen, D. (eds.) Information Processing in Medical Imaging
  - 25th International Conference, {IPMI} Proceedings. Lecture Notes in
  Computer Science, vol. 10265, pp. 597--609. Springer (2017),
  \url{https://doi.org/10.1007/978-3-319-59050-9\_47}

\bibitem{kamnitsas2017_deepmedic}
Kamnitsas, K., Ledig, C., Newcombe, V.F.J., Simpson, J.P., Kane, A.D., Menon,
  D.K., Rueckert, D., Glocker, B.: Efficient multi-scale 3d {CNN} with fully
  connected {CRF} for accurate brain lesion segmentation. Medical Image
  Analysis  \textbf{36},  61--78 (2017),
  \url{https://doi.org/10.1016/j.media.2016.10.004}

\bibitem{mousavi2017robust}
Mousavi, S.M., Naghsh, A., Manaf, A.A., Abu-Bakar, S.: A robust medical image
  watermarking against salt and pepper noise for brain {MRI} images. Multimedia
  Tools and Applications  \textbf{76}(7),  10313--10342 (2017)

\bibitem{osadebey2018blind}
Osadebey, M.E., Pedersen, M., Arnold, D.L., Wendel-Mitoraj, K.E.: Blind blur
  assessment of {MRI} images using parallel multiscale difference of gaussian
  filters. Biomedical engineering online  \textbf{17}(1),  1--22 (2018)

\bibitem{perone2019}
Perone, C.S., Ballester, P.L., Barros, R.C., Cohen{-}Adad, J.: Unsupervised
  domain adaptation for medical imaging segmentation with self-ensembling.
  NeuroImage  \textbf{194},  1--11 (2019),
  \url{https://doi.org/10.1016/j.neuroimage.2019.03.026}

\bibitem{scipy}
Virtanen, P., Gommers, R., Oliphant, T.E., Haberland, M., Reddy, T.,
  Cournapeau, D., Burovski, E., Peterson, P., Weckesser, W., Bright, J., {van
  der Walt}, S.J., Brett, M., Wilson, J., Millman, K.J., Mayorov, N., Nelson,
  A.R.J., Jones, E., Kern, R., Larson, E., Carey, C.J., Polat, {\.I}., Feng,
  Y., Moore, E.W., {VanderPlas}, J., Laxalde, D., Perktold, J., Cimrman, R.,
  Henriksen, I., Quintero, E.A., Harris, C.R., Archibald, A.M., Ribeiro, A.H.,
  Pedregosa, F., {van Mulbregt}, P., {SciPy 1.0 Contributors}: {{SciPy} 1.0:
  Fundamental Algorithms for Scientific Computing in Python}. Nature Methods
  \textbf{17},  261--272 (2020)

\bibitem{scikit-image}
van~der Walt, S., Sch{\"{o}}nberger, J.L., Nunez{-}Iglesias, J., Boulogne, F.,
  Warner, J.D., Yager, N., Gouillart, E., Yu, T.: scikit-image: Image
  processing in python. CoRR  \textbf{abs/1407.6245} (2014),
  \url{http://arxiv.org/abs/1407.6245}

\end{thebibliography}

\end{document}